\documentclass[aps,prl,reprint,floatfix]{revtex4-1}

\usepackage{natbib,graphicx,amssymb,amsmath,xcolor,bm,hyperref}

\begin{document}
\title[]{Analytic regimes of tokamak plasma}

\author{Alexander M. Balk}
\affiliation{Department of Mathematics, University of Utah,
155 South 1400 East, Salt Lake City, Utah 84112}

\date{\today}

\begin{abstract}
The paper suggests certain regimes of fusion plasma with magnetic confinement, when the Larmor radius $\lambda$ and the angle $\alpha$ (determining the orientation of the local coordinates) form an analytic function 
$w=\lambda e^{i\alpha}$ of complex variable $z=x_1+i x_2$ ($x_1$ and $x_2$ are coordinates of tokamak cross-section). It is known that an efficient nuclear fusion requires transport barriers, which might be due to zonal/poloidal flow; and the emergence of such flow is implied by the adiabatic extra conservation (established in a local slub). The suggested analytic regimes are based on the global extension of this extra conservation; transport barriers could spread to the entire tokamak plasma. It is interesting that in these regimes, plasma temperature reaches maximum often not at a single location (of tokamak cross-section), but at several points separated by regions of lower temperature. 
\end{abstract} 

\maketitle

In nuclear fusion devices with magnetic confinement --- such as tokamaks --- it is imortant to have transport barriers, which might be due to zonal/poloidal flow \cite{Horton, Di}. It is interesting that the emergence of such flow in a plasma slub is implied by the conservation of the adiabatic extra invariant, along with the energy and enstrophy. In the present paper, we study whether this invariant can be extended from the conservation in a local slub to the global conservation in the entire tokamak plasma. We find that such extension leads to certain analiticity regimes. 

{\it 1. Extra invariant.}
Suppose a significant mode of plasma dynamics is represented by the drift waves with dispersion relation 
$\omega_{\bf k}=q/(1+p^2+q^2)$. The interaction of these waves can take different forms. The dispersion is written in dimensionless form for a local slub: The local wave number components $p,q$ of the wave vector ${\bf k}$ are non-dimensionalized by the Larmor radius $\lambda$. 

The radial transport in a tokamak is reduced  when the wave system organizes itself to produce strong alternating zonal/poloidal flow \cite{BiglariDiTe}. This is because the ballistic transfer of heat and particles in the radial direction --- by {\it any} plasma mode --- is interrupted (radial velocity field is de-correlated) when there is strong randomly alternating poloidal current. At the same time, the poloidal flow does not contribute to radial transport. 

\begin{figure}
\includegraphics[width=\columnwidth]{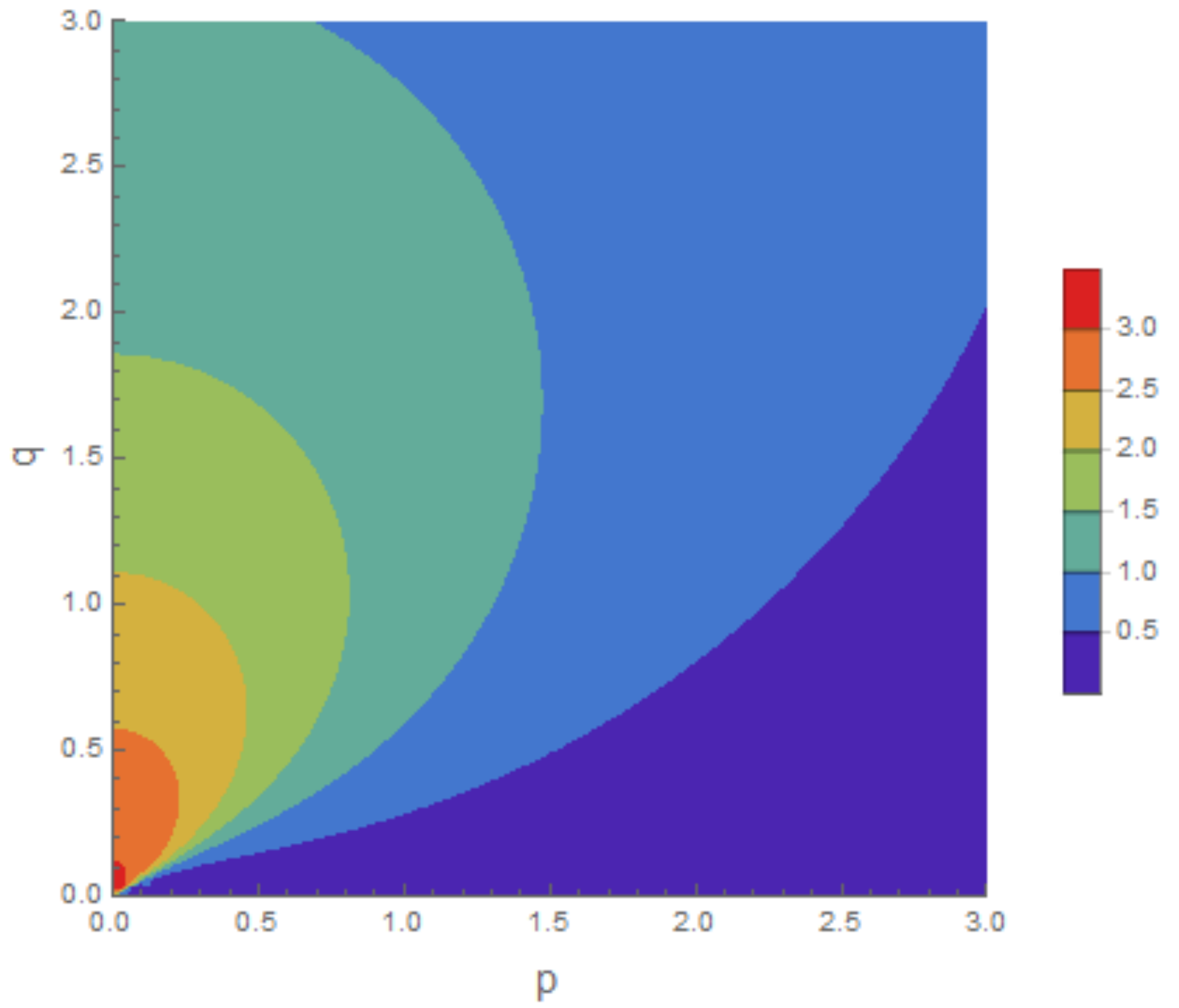}
\caption{Contour plot of the function (\ref{extraDnsty}). Color represents the values of $\eta(p,q)$. The graph is shown only for positive $p,q$, due to the symmetries $\eta(p,q)=\eta(-p,q)=-\eta(p,-q)$.
All contour lines pass through the origin with $30^\circ$ slope.}
\label{fig:invar}
\end{figure}

It is interesting that the emergence of zonal/poloidal flow follows from the extra conservation.
In addition to the energy and enstrophy, a system of drift waves is known \cite{BNZ,B1991} to conserve one more quadratic integral $I=\int \eta_{\bf k} {\mathcal N}_{\bf k}\, d{\bf k}$. This is the extra invariant
with spectral density
\begin{eqnarray}\label{extraDnsty}
\eta_{\bf k}=\arctan{\frac{p+q\sqrt{3}}{p^2+q^2}}-\arctan{\frac{p-q\sqrt{3}}{p^2+q^2}}\,.
\end{eqnarray}
The function ${\mathcal N}_{\bf k}=\varepsilon_{\bf k}/\omega_{\bf k}$ is the wave action spectrum, 
$\varepsilon_{\bf k}$ is the energy spectrum. 
The extra conservation is {\it adiabatic}: It holds approximately over long time. The function $\eta(p,q)$ is shown in Figure \ref{fig:invar}.

As well known, the conservation of the energy and enstrophy leads to the inverse energy cascade (local or non-local). The conservation of the extra invariant ensures \cite{B2005} that the energy is not just transferred towards small wave numbers, but concentrates around $p$-axis ($|q|\ll |p|$), which corresponds to zonal/poloidal flow. (In the last three references, $p$ and $q$ are interchanged for geophysical context.)

The waves similar to the plasma drift waves occur in different problems. As well known, the waves with the same dispersion relation are the Rossby waves in atmospheres and oceans. Much slower waves (slower than the relatively slow Rossby waves) also appear in astrophysical magnethydrodynamics \cite{Hide66,ZaqOliBalShe,B2022}. In all these situations, the extra invariant has important physical implications.

{\it 2. Global conservation.} 
Let $(x_1,x_2,x_3)$ be global dimensional coordinates; $x_3$ is the toroidal coordinate. In all cross-sections $x_3=const$, plasma properties are assumed to be the same.  
In the present paper, we assume that the confining magnetic field ${\bf B}$ is purely toroidal, ${\bf B}=B\hat x_3$; and so, it is orthogonal to the cross-section. 

The drift waves have the dispersion relation $\Omega=\nu(x_1,x_2)\, \omega(p,q)$,
where $p$ and $q$ are local dimensionless wave numbers; 
they are related to the global dimensional wave vector ${\bf k}=(k^1, k^2)$ 
by a rotation through angle $\alpha(x_1,x_2)$ $\quad[c\equiv\cos\alpha,\;s\equiv\sin\alpha]$
\begin{eqnarray}\label{pq}
p=\lambda(c\,k^1 + s\,k^2), \quad q=\lambda(-s\,k^1 + c\,k^2).
\end{eqnarray}

In the plane ${\bf x}=(x_1,x_2)$,
the drift wave field is described by the inhomogeneous wave kinetic equation
\begin{eqnarray}\label{WKE}
\frac{\partial {\mathcal N}}{\partial t}+
  \frac{\partial \omega}{\partial \bf k}\cdot\frac{\partial \mathcal N}{\partial \bf x}
-\frac{\partial \omega}{\partial \bf x}\cdot\frac{\partial \mathcal N}{\partial \bf k}=St[{\mathcal N}]
\end{eqnarray}
for the dynamics of the wave action spectrum ${\mathcal N}({\bf k},{\bf x}, t)$; $t$ is dimensionless time (nondimensionalized by the frequency $\nu$).
(The wave kinetic equation has been utilized for a long time in physics and applied mathematics;
recently, kinetic equation approach was used \cite{parker16,MarstonTobias22} to study a related problem: interaction between zonal flow and turbulence.) The l.h.s.\ in equation (\ref{WKE}) describes the slow refraction of waves, while the r.h.s. (stoss-term or collision integral) describes changes in $\mathcal N$ due to 3-wave resonance interactions
\begin{eqnarray}\label{Res}
{\bf k}_3={\bf k}_1+{\bf k}_2, \quad \omega({\bf k}_3)=\omega({\bf k}_1)+\omega({\bf k}_2).
\end{eqnarray}
We assume that the interactions occur locally in physical space, and so, the three interacting waves have the same plasma parameters.

{\it Note:} We can consider only {\it linear} equations but study {\it nonlinear} interactions. This is because --- in general --- the conservation of a quadratic integral is determined by the dispersion relation. 
This fact was noted by Zakharov and Schulman: They called {\it degenerative}  \cite{ZSch0} a dispersion relation that allows for an additional conservation law (besides the energy, momentum, and wave action). They showed \cite{ZSch} that conservation in the homogeneous wave kinetic equation is equivalent to cancellation of the ``small divisors'', and therefore, to approximate extra conservation in the dynamic equation. 
The extra conservation holds for {\it any nonlinearity},  provided the dynamic equation is {\it Hamiltonian} (which is usually the case for a physical system).
I believe the conservation by the inhomogenous part, on the left of the wave kinetic equation (\ref{WKE}), can also be studied in terms of the dynamic equation.

Let us check the global conservation of an arbitrary integral 
\begin{eqnarray}\label{I}
I({\bf x}, t)=\int F({\bf k},{\bf x})\,{\mathcal N}({\bf k},{\bf x},t) \, d{\bf k},
\end{eqnarray}
determined by an arbitrary spectral density $F$. According to the equation ({\ref{WKE}),
\begin{eqnarray}\label{D}
\frac{\partial I}{\partial t}&+&
\frac{\partial}{\partial {\bf x}}\cdot\int\frac{\partial\omega}{\partial {\bf k}}\,F\,{\mathcal N}\,d{\bf k}+
\int \frac{\partial (F,\omega)}{\partial({\bf k},{\bf x})}{\mathcal N}\,d{\bf k}\nonumber\\ 
&=&\int St[{\mathcal N}] \, F \, d{\bf k},
\end{eqnarray}
where we use notation
\begin{eqnarray*}
\frac{\partial ({\mathcal F},{\mathcal G})}{\partial(u,v)}=
\frac{\partial {\mathcal F}}{\partial u}\cdot \frac{\partial {\mathcal G}}{\partial v}\;-\;
\frac{\partial {\mathcal F}}{\partial v}\cdot \frac{\partial {\mathcal G}}{\partial u}
\end{eqnarray*}    
for arbitrary functions ${\mathcal F}$, ${\mathcal G}$ of arbitrary variables $u$, $v$ (vector or scalar). The quantity $I$ is conserved if in the equation (\ref{D}), the last two integrals vanish.

The wave kinetic equation (\ref{WKE}) always conserves the energy, when $F=\omega$. 

{\it 3. Anlyticity.} Now let $F$ be the function (\ref{extraDnsty}), $F=\eta$, with $p,q$ determined in (\ref{pq}). Then the r.h.s.\ in (\ref{D}) vanishes. Indeed
 by (\ref{pq}), $p$ and $q$ are linear combinations of the global components of the wave vector ${\bf k}$, and so, the equations (\ref{Res}) lead to
$p_3=p_1+p_2,\;  q_3=q_1+q_2,\; \omega(p_3,q_3)=\omega(p_1,q_1)+\omega(p_2,q_2).$
These equations imply that
\begin{eqnarray}\label{conservedF}
\eta(p_3,q_3)=\eta(p_1,q_1)+\eta(p_2,q_2),
\end{eqnarray}
i.e. the function $\eta$ is conserved in 3-wave resonance interactions. 
Actually, $\eta$ is the only independent function conserved in the 3-wave interactions. This fact is valid even if the equation (\ref{conservedF}) holds only in an arbitrarily small neighborhood anywhere on the resonance manifold \cite{Ferapontov,BaFe}.
All other conserved functions are linear combinations of $p$, $q$, $\omega(p,q)$, and $\eta(p,q)$.

 In the second integral of equation (\ref{D}),
\begin{eqnarray}\label{Jac1}
\frac{\partial (F,\omega)}{\partial({\bf k},{\bf x})}&=&
\frac{\partial (\eta,\omega)}{\partial(p,q)}
\frac{\partial (p,q)}{\partial({\bf k},{\bf x})}.
\end{eqnarray}
According to the equations (\ref{pq}),
\begin{eqnarray}\label{Jac2}
\frac{\partial (p,q)}{\partial({\bf k},{\bf x})}=
\lambda{\bf k}\cdot 
\left(-\frac{\partial\lambda}{\partial x_2}-\lambda\frac{\partial\alpha}{\partial x_1},\;
\frac{\partial\lambda}{\partial x_1}-\lambda\frac{\partial\alpha}{\partial x_2}\right).
\end{eqnarray}
So, the expressions (\ref{Jac2}) and (\ref{Jac1}) vanish for all wave vectors ${\bf k}$ if functions $\mu=\ln\lambda$ and $\alpha$ satisfy the Cauchy-Riemann conditions:
\begin{eqnarray}\label{Anali}
\frac{1}{\lambda}\frac{\partial\lambda}{\partial x_1}=\frac{\partial\alpha}{\partial x_2},\qquad
\frac{1}{\lambda}\frac{\partial\lambda}{\partial x_2}=-\frac{\partial\alpha}{\partial x_1},
\end{eqnarray}
saying that $\mu+i\alpha$ is an analytic function of $z=x_1+ix_2$. In other words, $w=\lambda e^{i\alpha}$ should be an analytic function of $z$.

The drift waves are determined by the gradient of some macroscopic plasma quantity. When this quantity reaches maximum, the frequency $\nu$ vanishes. There are no drift waves at such locations, the entire reasoning based on the wave kinetic equation (\ref{WKE}) fails, and the analyticity is not required. The analytic function $w(z)$ could have singularities at such points.

{\it 4. Examples.} 
Since $\lambda$ is related to the plasma temperature (e.g. $\lambda\propto\sqrt{T}$, with $T$ being the ion temperature), we would like to have smaller $\lambda$ at the tokamak boundary and larger $\lambda$ in the central part of the tokamak. The figures below show the Larmor radius $\lambda=|w(z)|$ in the logarithmic scale: The color bar numbers are the values of $\ln\lambda(x_1,x_2)$; the white area corresponds to the values exceeding the top color.

Example 1. 
\begin{eqnarray}\label{2}
w=\frac{A_1}{z-a_1}+\frac{A_2}{z-a_2} 
\end{eqnarray}
with four complex parameters $a_1, a_2, A_1, A_2$.
To ensure fastest decrease of $\lambda=|w|$ as $z\rightarrow\infty$, 
we set $A_1+A_2=0$ so that $w=O(z^{-2})$.
Using rotation and stretching in the $w$-plane, we can assume $A_1=-A_2=1$. Choosing the axis $x_1,\, x_2$, we can have $a_1=-a_2=i b$ ($b$ is a real number). Re-scaling length, we set $b=1$. See figure \ref{fig:two}.
\begin{figure}[h!]
\includegraphics[width=\columnwidth]{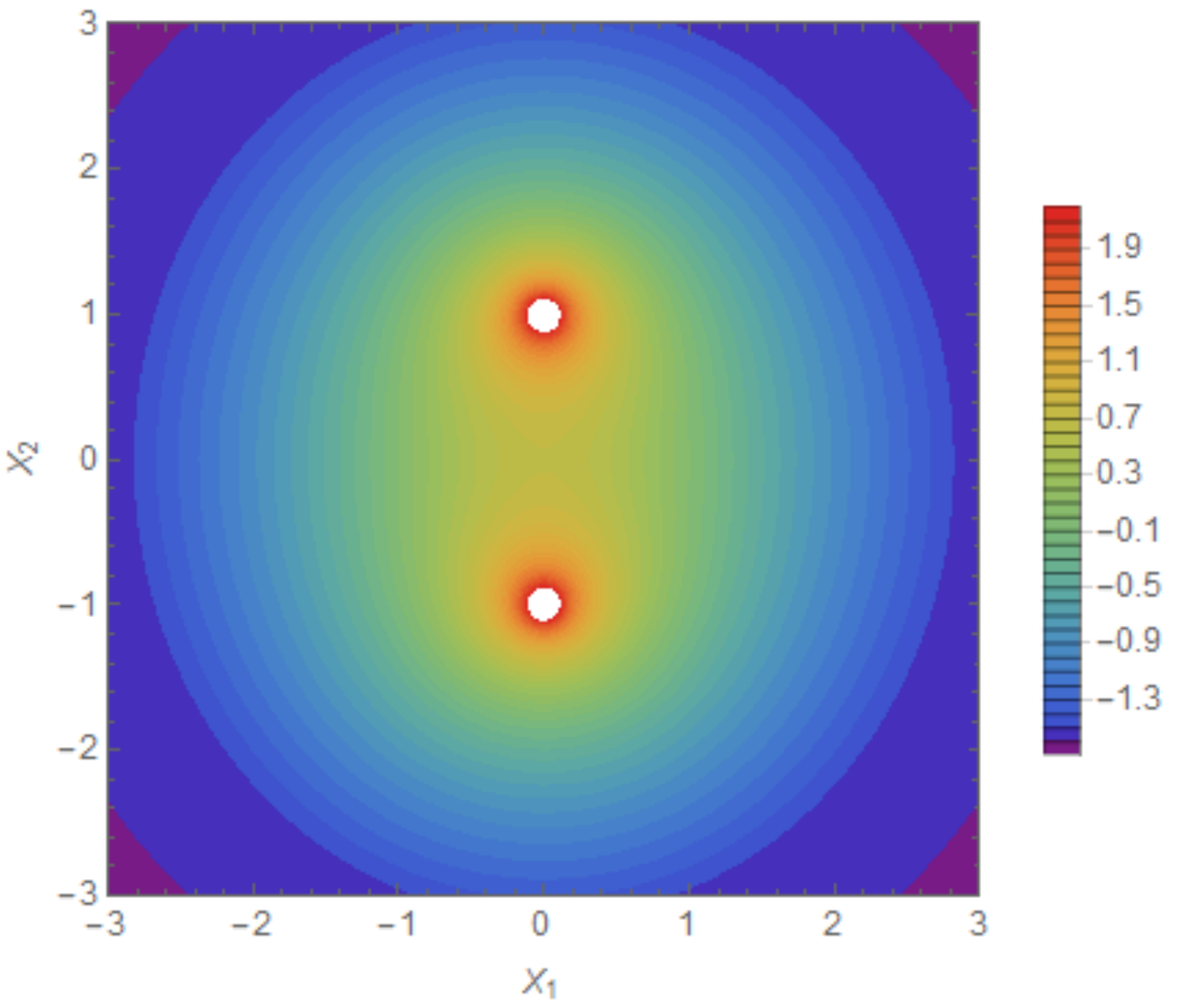}
\caption{
$w=\frac{1}{z-i}-\frac{1}{z+i}$.}
\label{fig:two}
\end{figure}
The boundary of the tokamak cross-section should pass through the region with smaller values of $\lambda$.
 
Example 2. 
\begin{eqnarray}\label{3}
w=\frac{A_1}{z-a_1}+\frac{A_2}{z-a_2}+\frac{A_3}{z-a_3}
\end{eqnarray}
with six complex parameters $A_j, a_j \; (j=1, 2, 3)$.
Similar to example 1, we require  $w=O(z^{-3})$, which gives us the two linear equations on $A_1, A_2, A_3$. Using rotation and stretching in the $w$-plane, we can set 
\begin{eqnarray}\label{Aa}
A_1=a_3-a_2,\; A_2=a_1-a_3,\; A_3=a_2-a_1. 
\end{eqnarray}
Finally, using rotation, shifting, and stretching in the $z$-plane, we can exclude 4 real parameters, leaving us with two parameters. See figure \ref{fig:three}.
\begin{figure}[h!]
\includegraphics[width=\columnwidth]{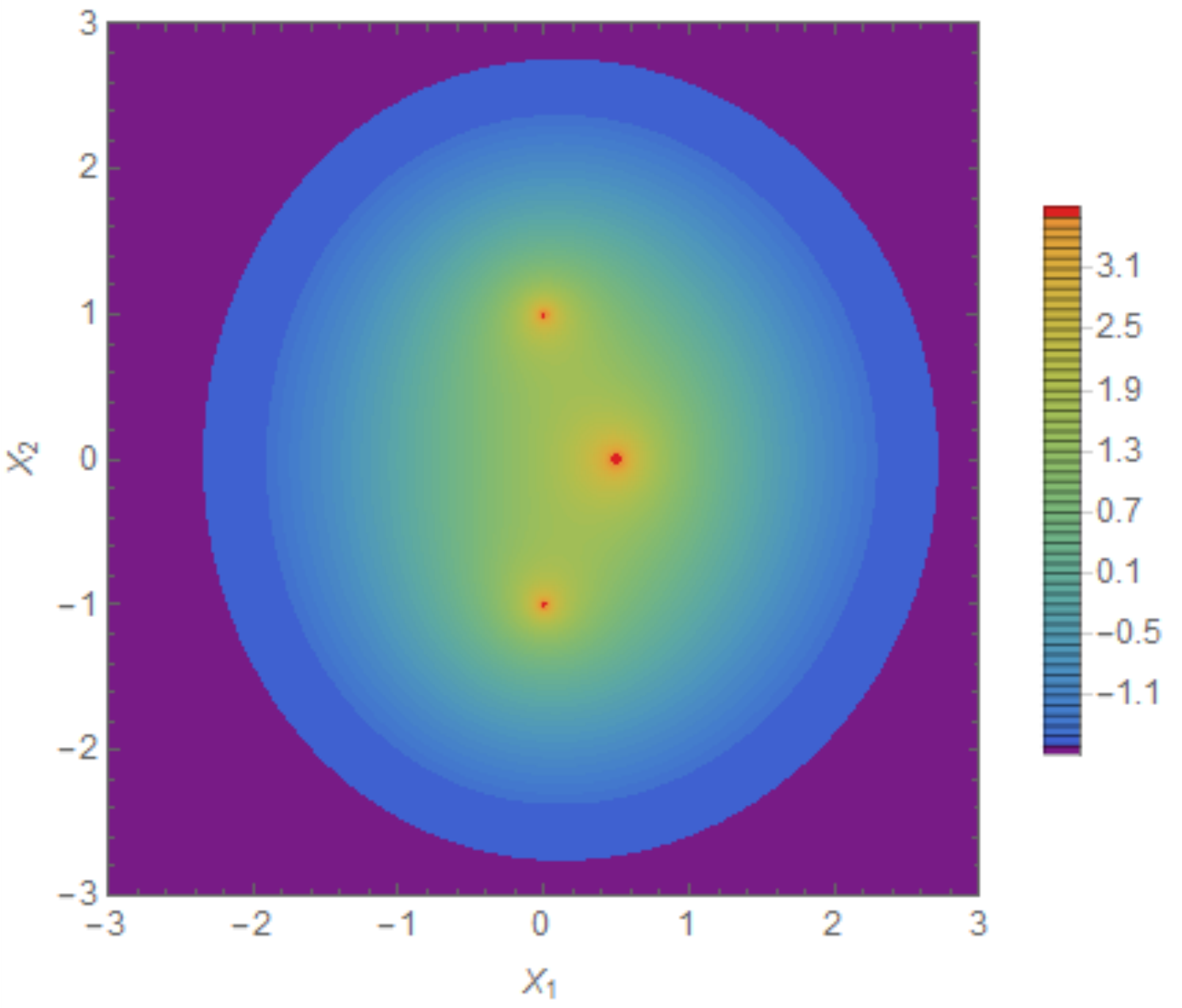}
\caption{The analytic function $w(z)$ is defined in (\ref{3})-(\ref{Aa}) with $a_1=i,\, a_2=-i,\, a_3=1/2$.}
\label{fig:three}
\end{figure}

\bibliography{My}
\end{document}